\documentclass[prd,aps,twocolumn]{revtex4} 

\usepackage{graphicx}
\usepackage{dcolumn}
\usepackage{bm}

\newcommand{\plb}[2]{{\em Phys. Lett.}              {\bf B#1}, #2 }
\newcommand{\npb}[2]{{\em Nucl. Phys.}              {\bf B#1}, #2 }

\newcommand{\prt}[2]{{\em Phys. Rev.}               {\bf D#1}, #2 }
\newcommand{\pru}[2]{{\em Phys. Rev. Lett.}         {\bf  #1}:#2 }

\newcommand{\be}{\begin{equation}}
\newcommand{\ee}{\end{equation}}
\newcommand{\ba}{\begin{eqnarray}}
\newcommand{\ea}{\end{eqnarray}}
\newcommand{\bt}{\begin{table}}
\newcommand{\et}{\end{table}}
\newcommand{\brt}{\begin{ruledtabular}}
\newcommand{\ert}{\end{ruledtabular}}
\newcommand{\btu}{\begin{tabular}}
\newcommand{\etu}{\end{tabular}}

\def\tr{{\rm Tr}}

\def\al2{\frac{\alpha^2}{\pi^2}}

\def\al{\alpha}
\def\th{\theta}
\def\ka{\kappa}
\def\la{\lambda}
\def\lt{\left}
\def\rt{\right}
\def\ov{\over}
\def\nonu{\nonumber}

\begin{document}

\preprint{
\noindent
\hfill
\begin{minipage}[t]{3in}
\begin{flushright}
\vspace*{2cm}
\end{flushright}
\end{minipage}
}


\title{Two-loop Renormalization Group Equations in General Gauge Field Theories}
\author{Mingxing Luo, Huawen Wang and Yong Xiao}
\affiliation{Zhejiang Institute of Modern Physics, Department of Physics,
Zhejiang University, Hangzhou, Zhejiang 310027, P R China}

\date{\today}

\begin{abstract}
The complete set of two-loop renormalization group equations in general gauge 
field theories is presented.
This includes the $\beta$-functions of parameters with and without a mass
dimension.
\end{abstract}


\maketitle

\section{Introduction}

Renormalization group equations (RGEs) provide a unique means in the analysis
of particle physics.
Comprehensive analysis of RGEs confirmed the behavior of asymptotic freedom in QCD,
which played a pivotal role in establishing 
a non-Abelian gauge theory for the strong interaction \cite{pdg}.
The runnings of coupling constants and mass parameters are crucial in global analysis of
high precision electroweak experiments \cite{erler}.
On the other hand, RGEs analysis extrapolated to extremely high energy provides a
possible, in some cases the only feasible, test for physics beyond the SM.
The RGEs is a natural ingredient in the analysis of  grand unification theories and string theories.
For more than ten year, it has been know that gauge couplings do not unify within the SM.
This gives extra evidence against simple grand unification theories such as $SU(5)$ without supersymmetry,
in addition to the non-observation of proton decay.
On the other hand, gauge couplings seem to unify
at a scale $\sim 2\times 10^{16} \ GeV$ in the minimal
supersymmetric standard model, which can be interpreted
as an indirect evidence for supersymmetry as well as
unification theories \cite{GUT1,GUT2,GUT3}.
 
Computations of RGEs in gauge theories have been performed for various models
to different orders of perturbation.
Persistent efforts yielded recently a four-loop result of the
$\beta$-function of the strong coupling constant \cite{4loops}.
Two-loop RGEs of dimensionless couplings in a general gauge theory
as well as the specific case of the SM had been
calculated long ago in a series
of classic papers by Machacek and Vaughn \cite{MV1,MV2,MV3}.
By introducing a non-propagating gauge-singlet ``dummy" scalar field,
two-loop RGEs of dimensional couplings
can be readily inferred from dimensionless results \cite{martin,luo}.
These were used to derive the RGEs of supersymmetric theories
a decade later \cite{martin}.

Recently\cite{luo}, we re-calculated the two-loop RGEs in the SM, in a combination
of using the general results of \cite{MV1,MV2,MV3} and direct calculations from
Feynman diagrams.
A new coefficient was found in the $\beta$-function of the quartic coupling
and a class of gauge invariants were found to be absent in $\beta$-functions of
hadronic Yukawa couplings.
We also presented the two-loop $\beta$-function of the Higgs mass parameter
in complete form, which provided a partial but useful check on
the calculation of the quartic coupling.
Whenever discrepancy with the literature appears, we carefully inspected
relevant Feynman diagrams to ensure consistency. 
In this paper, we present
the complete set of two-loop renormalization group equations in general gauge field theories.
This includes the $\beta$-functions of parameters with and without a mass dimension.
The results in \cite{luo} can then be readily reproduced.

In section 2, we present essential notations and definitions, along with a discussion
of the differences between \cite{MV1,MV2,MV3} and our analysis.
In section 3, we present the $\gamma$-functions of the scalar and fermion fields.
In section 4, we present the $\beta$-functions of dimensionless parameters and
in section 5 those of dimensional parameters. 
In section 6 the results are extended to semi-simple groups.
We conclude in section 7.

\section{Notations and Definitions}
We start with a general renormalizable field theory with gauge fields
$V_\mu^A$ of a compact simple group $G$, scalar fields $\phi_a$ and two-component 
fermion fields $\psi_j$. 
In section 6, these results will be extended to semi-simple groups. 
The Lagrangian of the theory can be conveniently divided into three parts,
\ba
{\cal L} = {\cal L}_0 + {\cal L}_1
           + \lt( {\rm gauge \ fixing + ghost \ terms} \rt),
\label{lagragian}
\ea
where ${\cal L}_0$ contains no dimensional parameters and ${\cal L}_1$ includes all
terms with dimensional parameters.
Explicitly
\ba
{\cal L}_0 &= & - {1\ov4} F^{\mu\nu}_A F_{\mu\nu}^A + {1\ov2}D^\mu \phi_a  D_\mu \phi_a
           + i\psi_j^+ \sigma^\mu D_\mu \psi_j  \nonu \\  
         &-&   {1\ov2} \lt( Y_{jk}^a \psi_j \zeta \psi_k \phi_a + h.c. \rt) 
           - {1\ov4!} \la_{abcd} \phi_a \phi_b \phi_c \phi_d,
\label{lag1}
\ea
where $\zeta=\pm i \sigma_2$.
Different from \cite{MV1,MV2,MV3}, 
we have included an overall ${1\ov2}$ factor in the Yukawa coupling terms.
The gauge field strengths are defined to be
\be
F_{\mu\nu}^A = \partial_\mu V_\nu^A - \partial_\nu V_\mu^A 
             + g f^{ABC}  V_\mu^B V_\nu^C
\ee
where $f^{ABC}$ are the structure constants of the gauge group
and $g$ is the gauge coupling constant.
Choosing the standard $R_\xi$ gauge, the gauge field propagator is
\be
D_{\mu\nu}^{AB}(k) = \delta^{AB} \left( - g_{\mu\nu} + (1-\xi) {k_\mu k_\nu \over k^2} \right)
{i \over k^2},
\ee
where $\xi$ is the gauge parameter. Covariant derivatives of matter fields are
\ba
D_\mu \phi_a = \partial_\mu \phi_a - ig\th_{ab}^A V_\mu^A \phi_b \\
D_\mu \psi_j = \partial_\mu \psi_j - igt_{jk}^A V_\mu^A \psi_k
\ea
where both $\th_{ab}^A$ and $t_{jk}^A$ are hermitian matrices, which form 
representations of the gauge group on the scalar and fermion fields, respectively.
Since any complex scalar fields can always be decomposed in terms of real ones,
scalars in this paper are assumed to be real.
$\th^A$ are thus pure imaginary and antisymmetric. 
For late convenience, we define the following gauge invariants:
\ba
C_2^{ab} (S) = \th^A_{ac} \th^A_{cb}, \ S_2(S) \delta^{AB} = \tr [\th^A \th^B] \\
C_2^{ab} (F) = t^A_{ac} t^A_{cb}, \ S_2(F) \delta^{AB} = \tr [t^A t^B] \\
C_2(G) \delta^{AB} = f^{ACD} f^{BCD}  
\ea
$C_2^{ab}(R)$ is block diagonal for each irreducible representation $R(\ =S,F)$ of
eigenvalue of $C_2(R)$.

In this paper, we use dimensional regularization and the modified minimal subtraction algorithm.
The renormalized coupling constants $x_k$ in 
$d=4-2 \epsilon$ are related to corresponding bare coupling constants 
$x_k^0$ by 
\be
 x_k^0 \mu^{- \rho_k \epsilon} = x_k + \sum_{n=1}^{\infty} a_k^{(n)} {1\over\epsilon^n}
\ee
where $\mu$ is an arbitrary mass scale parameter, 
$\rho_k=1(2)$ for gauge and Yukawa (scalar quartic) coupling constants, 
and $ a_k^{(n)} $ are to be calculated perturbatively.
The $\beta$-functions of  $x_k$  are defined to be
\be
\beta_{x_k} = \left. \mu {d x_k \over d \mu} \right|_{\epsilon=0}   
\ee
It is easy to see that
\be
\beta_{x_k} = \sum_l \rho_l x_l {\partial a_k^{(1)} \over \partial x_l}
             - \rho_k a_k^{(1)}
\ee
Perturbatively, one has
\be
\beta_{x_k} = {1 \over (4 \pi)^2} \beta_{x_k}^{I} + {1 \over (4 \pi)^4} \beta_{x_k}^{II} + \cdots
\ee
where $\beta_{x_k}^{I}$ and $\beta_{x_k}^{II}$ are one- and two-loop contributions, respectively.
The wave function renormalization constant $Z_i$ of the $i$-th field can be expressed as 
\be
Z_i = 1 + \sum_{n=1}^{\infty} C_i^{(n)} {1\over\epsilon^n} 
\ee
The corresponding anomalous dimension is 
\be
\gamma_i = {1\over2} \mu {d \over d \mu} \log Z_i
 = - {1\over2}  \sum_l \rho_l x_l {\partial C_i^{(1)} \over \partial x_l}.
\ee
Also perturbatively, one has
\be
\gamma_{i} = {1 \over (4 \pi)^2} \gamma_i^{I} + {1 \over (4 \pi)^4} \gamma_{i}^{II} + \cdots 
\ee
where $\gamma_i^{I}$ and $\gamma_{i}^{II}$ are one- and two-loop contributions, respectively.

The constraint on Yukawa coupling matrices
imposed by gauge invariance is given by
 \be
 Y_{jk}^b \theta_{ba}^A + Y_{jl}^a t_{lk}^A + t_{jl}^{A*} Y_{lk}^a = 0.
\label{yukawa}
 \ee
When a fermion is involved simultaneously with a Yukawa coupling and a gauge coupling, 
one usually has the combination $Y_{jl}^a t_{lk}^A + t_{jl}^{A*} Y_{lk}^a$.
This is to say, in these cases,
$Y^a$ is always preceded by a $t^{A*}$ or followed by a $t^A$.  
Eq. (\ref{yukawa}) can thus be used to simplify the gauge structure of Feynman diagrams.
In \cite{MV1,MV2,MV3}, fermion fields are implicitly assumed to be real
so $t^A$ are pure imaginary and antisymmetric. $t_{jl}^{A\star}$ was substituted by
$ - t_{jl}^{A}$ and eq.~ (\ref{yukawa}) is reduced to
 \be
 Y_{jk}^b \theta_{ba}^A + Y_{jl}^a t_{lk}^A - t_{jl}^A Y_{lk}^a = 0.
 \label{MVyukawa}
 \ee
However, in most physical theories, fermions are in general Weyl and complex, 
with only possible exceptions of neutrinos and gauginos as real Majorona fermions. 
Fortunately, close inspections of Feynman diagrams show that 
both approaches yield the same final results in most cases.
In cases where they differ, we will retain $t_{jl}^{A\star}$.

\begin{figure}
\includegraphics[width=3.5in]{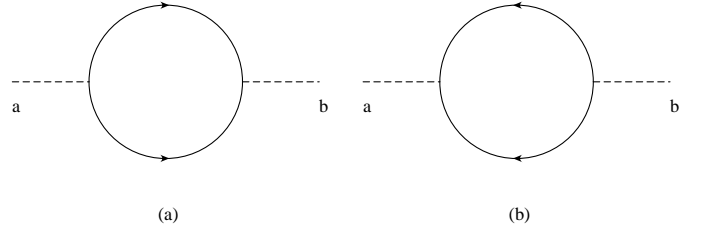}
\caption{\label{fig:Y2F}Fermion radiative corrections to scalar propagator}
\end{figure}

By assuming real fermion fields,
the direction of a fermion propagator needs not to be discriminated either.
Shown in Fig.~\ref{fig:Y2F} are two
fermion loop diagrams which contribute to the propagators of scalar fields.
The two diagrams are different for complex fermions and the total result is proportional to 
\be
 Y_2^{ab}(S) = {1\over2} \tr (Y^{+a} Y^b + Y^{+b} Y^a).
\label{Y2F}
\ee
$Y_2^{ab}(S)$ forms a hermitian matrices and 
is block diagonal for each irreducible representation of eigenvalue $Y_2(S)$.
For real fermions, the two terms in $Y_2^{ab}(S)$ become equal and they reduce to
a common factor $\tr (Y^{+a} Y^b)$.
Again, we will retain eq. (\ref{Y2F}) and other similar combinations.

The second part of the Lagrangian is,
\ba
{\cal L}_1 &=& - {1\ov2} \lt[ (m_f)_{jk} \psi_j \zeta \psi_k + h.c. \rt] 
     -  {{m_{ab}^2}\ov{2!}} \phi_a \phi_b  \nonumber \\
     &&          - {{h_{abc}}\ov{3!}} \phi_a \phi_b \phi_c 
\ea          
In principle, the $\beta$ functions of $(m_f)_{jk}$, $m_{ab}^2$ and $h_{abc}$ can
be calculated directly. But this would be tedious and can be avoided.
To do so, 
we introduce a non-propagating dummy real scalar field $\phi_{\hat{d}}$ with
no gauge interactions and rewrite the Lagrangian as,
\ba
{\cal L}_1 &=& 
        - {1\ov2} \lt( Y_{jk}^{\hat{d}} \psi_j \zeta \psi_k \phi_{\hat{d}} + h.c \rt)
           - {{\la_{ab\hat{d}\hat{d}}}\ov{4!}} \phi_a \phi_b \phi_{\hat{d}} \phi_{\hat{d}}
              \nonumber \\    
      &&     - {{\la_{abc\hat{d}}}\ov{4!}} \phi_a \phi_b \phi_c \phi_{\hat{d}}
\ea
with the substitution of 
$ Y_{ij}^{\hat{d}} = (m_f)_{ij} $, $ \la_{ab\hat{d}\hat{d}} = 2 m_{ab}^2 $ and
$ \la_{abc\hat{d}} = h_{abc} $. 
The $\beta$ functions of $m_f$, $m_{ab}^2$ and $h_{abc}$ are equal to those of
the new Yukawa coupling $Y^{\hat{d}}$, quartic scalar coupling
$\la_{ab\hat{d}\hat{d}}$,  and $\la_{abc\hat{d}}$.
The latter can be readily obtained from the 
$\beta$ functions of $Y^a$ and $\la_{abcd}$ by suppressing both the 
summations of the $\hat{d}$-type indices and related gauge couplings.

\section{Wave function renormalization}
\subsection{Scalar wave function renormalization}
To one loop, the anomalous dimensions of scalars are
\be
 \gamma_{ab}^{sI} = 2 \ka Y_2^{ab}(S) - g^2 (3-\xi) C_2^{ab}(S)
\ee
To two loops,
\ba
 \gamma_{ab}^{sII} &=&
- g^4 C_2^{ab}(S) \lt[ \lt( {35\ov3} - 2\xi - {1\ov4}\xi^2 \rt) C_2(G) \rt. \nonu \\ 
 && \ \ - \lt. {10\ov3} \ka S_2(F) - {11\ov12} S_2(S) \rt]  + {1\ov2} \Lambda_{ab}^2(S) \nonu \\
&+& {3\ov2}g^4 C_2^{ac}(S) C_2^{cb}(S) - 3 \ka  H_{ab}^2(S)  \nonu \\
&-& 2 \ka \bar{H}_{ab}^2(S)  + 10 \ka g^2 Y_{ab}^{2F}(S) 
\label{scalar2}
\ea
where $\ka=1/2(1)$ for two(four)-component fermions, and 
\ba
\Lambda_{ab}^2(S) &=& {1\ov6}\la_{acde}\la_{bcde} \\
 H_{ab}^2(S) &=& {1\ov2} \tr (Y^a Y^{+b} Y^c Y^{+c} + Y^{+a} Y^b Y^{+c} Y^c)  \\
\bar{H}_{ab}^2(S) &=& {1\ov2} \tr (Y^a Y^{+c} Y^b Y^{+c} + Y^{+a} Y^c Y^{+b} Y^c) \\
Y_{ab}^{2F}(S) &=& {1\ov2} \tr \lt[C_2(F)(Y^a Y^{+b} + Y^b Y^{+a}) \rt]
\ea
In contrast with \cite{MV1}, $Y_2^{ab}(S), \ H_{ab}^2(S), \bar{H}_{ab}^2(S)$ and
$Y_{ab}^{2F}(S)$ are modified so they are hermitian for complex fermions.
As mentioned above, it is due to that complex fermion lines 
have two distinct directions in Feynman diagrams.
The contributions of the two parts are hermitian conjugate to each other. 
For real fermions, both contributions are real.
So they are equal and can be combined.
Similar structure appears in later results.
Here and hereafter, we express the Casmir factors as
$C_2^{ab}(S)$ instead of $C_2(S) \delta_{ab}$
and $C_2^{ac}(S) C_2^{cb}(S)$ instead of $C_2^2(S)\delta_{ab}$,
to accommodate reducible representations.

\subsection{Fermion wave function renormalization}
To one loop, the anomalous dimensions of fermions are
\be
 \gamma^F_{I} = {1\ov2} Y^a Y^{+a} + g^2 C_2(F) \xi
\ee
To two loops,
\ba
 \gamma^F_{II} & = &
- {1\ov8} Y^a Y^{+b} Y^b Y^{+a} - {3\ov2} \ka Y^a Y^{+b} Y_2^{ab}(S) \nonu \\
&+& g^2 \lt[ {9\ov2} C_2^{ab}(S) Y^a Y^{+b} - {7\ov4} C_2(F) Y^a Y^{+a} \rt.  \nonu \\ 
  && \ - \lt.        {1\ov4} Y^a C_2(F) Y^{+a} \rt] \nonu \\
&+& g^4 C_2(F) \lt[ \lt({25\ov4} + 2\xi + {1\ov4}\xi^2 \rt) C_2(G)  \rt.  \nonu \\ 
  && \  - \lt. 2 \ka S_2(F)  - {1\ov4} S_2(S) \rt] - {3\ov2} g^4 \lt[ C_2(F) \rt]^2
\label{fermi2}
\ea
Here the only modification is in the $Y_2^{ab}(S)$, which appears in the first line of 
eq.(\ref{fermi2}). 
This is due to one-loop fermion sub-diagrams in scalar field propagators.

\section{Dimensionless parameters}
\subsection{The gauge coupling constant}
The $\beta$-function for the gauge coupling constants has been extensively studied.
Recent results are up to the fourth order. 
Here we include the two loop results for completeness,
\ba
\beta (g) &=& - {g^3\ov{(4\pi)^2}} \lt\{ {11\ov3} C_2(G) - {4\ov3}\ka S_2(F) - {1\ov6} S_2(S)
          \rt.     \nonu \\  
         &&\ \ \lt. + {{2\ka}\ov{(4\pi)^2}} Y_4(F) \rt\} - {{g^5}\ov{(4\pi)^4}} \lt\{ {34\ov3} \lt[ C_2(G) \rt]^2
          \rt.     \nonu \\  
         &&\ \ \ \ \lt. - \ka \lt[ 4 C_2(F) + {20\ov3} C_2(G) \rt] S_2(F) \rt. \nonu \\
         &&\ \ \ \  - \lt. \lt[ 2 C_2(S) + {1\ov3} C_2(G) \rt] S_2(S) \rt\}  \label{gague2}
\ea
where $Y_4(F)$ is defined through
\be
Y_4(F) = {1\ov{d(G)}} \tr \lt[ C_2(F) Y^a Y^{+a} \rt]
\ee
$d(G)$ is the dimension of the group.
Note that though the Yukawa couplings are normalized differently, 
eq. (\ref{gague2}) assumes the same form as eq (6.1) of \cite{MV1}.

\subsection{The Yukawa couplings}
The $\beta$-functions of the Yukawa couplings can be expressed as
\be
\beta^a = \gamma^a + \gamma^{+F} Y^a + Y^a \gamma^F + \gamma_{ab}^S Y^b
\ee
where $\gamma^a$ are the anomalous dimensions of the operators $\phi_a \psi_j \zeta \psi_k$,
$ \gamma^F$ and $\gamma_{ab}^S$ are the anomalous dimensions of the corresponding fermions 
and bosons, respectively.
To one loop,
\ba
 \beta^a_{I}&=&
 {1\ov2} \lt[Y_2^+(F) Y^a + Y^a Y_2(F) \rt] + 2 Y^b Y^{+a} Y^b \nonu \\
&& +  2 \ka Y^b Y_2^{ab}(S) - 3 g^2 \{ C_2(F), Y^a \}
\ea
where
\be
Y_2(F) = Y^{+a} Y^a
\ee
To two loops,
\ba
 \beta^a_{II} &=& 2 Y^c Y^{+b} Y^a (Y^{+c} Y^b - Y^{+b} Y^c) \nonu \\ 
&-& Y^b \lt[Y_2(F) Y^{+a} + Y^{+a} Y_2^+(F) \rt] Y^b \nonu \\
&-& {1\ov8} \lt[Y^b Y_2(F) Y^{+b} Y^a + Y^a Y^{+b} Y_2^+(F) Y^b \rt] \nonu \\
&-& 4\ka Y_2^{ac}(S) Y^b Y^{+c} Y^b - 2 \ka Y^b \bar{H}^2_{ab}(S) \nonu \\
&-& {3\ov2}\ka Y_2^{bc}(S) (Y^b Y^{+c} Y^a + Y^a Y^{+c} Y^b ) \nonu \\
&-& 3\ka Y^b H_{ab}^2(S)- 2 \la_{abcd} Y^b Y^{+c} Y^d + {1\ov2} \Lambda_{ab}^2(S) Y^b \nonu \\ 
&+& 3 g^2 \{ C_2(F), Y^b Y^{+a} Y^b \} + 5 g^2 Y^b \{ C_2(F), Y^{+a} \} Y^b \nonu \\ 
&-& {7\ov4} g^2 [C_2(F) Y_2^+(F) Y^a + Y^a Y_2(F) C_2(F)] \nonu \\
&-& {1\ov4} g^2 [Y^b C_2(F) Y^{+b} Y^a + Y^a Y^{+b} C_2(F) Y^b] \nonu \\
&+& 6 g^2 H_{2t}^a + 10 \ka g^2 Y^b Y_{ab}^{2F}(S) \nonu \\
&+& 6 g^2 [C_2^{bc}(S) Y^b Y^{+a} Y^c - 2 C_2^{ac}(S) Y^b Y^{+c} Y^b] \nonu \\
&+& {9\ov2} g^2 C_2^{bc}(S) (Y^b Y^{+c} Y^a + Y^a Y^{+c} Y^b) \nonu \\ 
&-& {3\ov2} g^4 \{ \lt[ C_2(F) \rt]^2, Y^a \} + 6g^4 C_2^{ab}(S) \{ C_2(F), Y^b \} \nonu \\ 
&+& g^4 \lt[ - {97\ov6}C_2(G) + {10\ov3}\ka S_2(F) \rt. \nonu \\ 
&+& \lt. {11\ov12}S_2(S) \rt] \{ C_2(F), Y^a \} - {21\ov2}g^4 C_2^{ab}(S) C_2^{bc}(S) Y^c \nonu \\ 
&+& g^4 C_2^{ab}(S) \lt[ {49\ov4} C_2(G) - 2\ka S_2(F) - {1\ov4}S_2(S) \rt] Y^b
\ea
where
\be
H_{2t}^a = t^{A*} Y^a Y^{+b} t^{A*} Y^b + Y^b t^A Y^{+b} Y^a t^A
\ee
The second term in the fourth line, the first term in the sixth line, and the second term in the tenth line, 
are expressed in terms of the hermitian gauge invariants 
$\bar{H}^2_{ab}(S)$, $H_{ab}^2(S)$, and $Y_{ab}^{2F}(S)$, respectively.
They are all from the $\gamma$-functions of the scalar fields. 
They differ from the fourth line and the ninth line of (3.3) in \cite{MV2},
which included only parts of the expressions. 
In $H_{2t}^a$, in front of $Y^a$ and $Y^b$ one has $t^{A*}$ instead of $t^A$, again due to the 
fact that fermions are complex.

\subsection{The scalar quartic couplings}
The $\beta$-functions of the scalar quartic couplings can be expressed as
\be
\beta_{abcd} = \gamma_{abcd} + \sum_i \gamma^S(i) \la_{abcd}
\ee
where $\gamma_{abcd}$ are the anomalous dimensions of the operators 
$\phi_a \phi_b \phi_c \phi_d$,
$ \gamma^S(i)$ is the anomalous dimension of the scalar field $i$.
To one-loop,
\ba
 \beta_{abcd}^{I} & = &
\Lambda_{abcd}^2 - 8\ka H_{abcd} + 2\ka \Lambda_{abcd}^Y \nonu \\ 
&& - 3g^2 \Lambda_{abcd}^S + 3g^4 A_{abcd}
\ea
where
\ba
\Lambda_{abcd}^2 &=& {1\ov8} \sum_{perms} \la_{abef} \la_{efcd} \\
H_{abcd} &=& {1\ov4} \sum_{perms} \tr( Y^a Y^{+b} Y^c Y^{+d} ) \\
\Lambda_{abcd}^Y &=& \sum_i Y_2(i) \la_{abcd} \\
\Lambda_{abcd}^S &=& \sum_i C_2(i) \la_{abcd} \\
A_{abcd} &=& {1\ov8} \sum_{perms} \{ \th^A, \th^B \}_{ab} \{ \th^A, \th^B \}_{cd} 
\ea
To two loops,
\ba
 \beta_{abcd}^{II} &=& {1\ov2} \sum_i \Lambda^2(i) \la_{abcd} - \bar{\Lambda}_{abcd}^3
   - 4\ka \bar{\Lambda}_{abcd}^{2Y} \nonu \\
&+& \ka \lt\{ 8\bar{H}_{abcd}^\la - \sum_i \lt[ 3H^2(i) + 2\bar{H}^2(i) \rt]
    \la_{abcd} \rt\} \nonu \\
&+& 4\ka(H_{abcd}^Y + 2\bar{H}_{abcd}^Y + 2H_{abcd}^3) \nonu \\
&+& g^2 \lt\{ 2\bar{\Lambda}_{abcd}^{2S} - 6\Lambda_{abcd}^{2g} + 4\ka(H_{abcd}^S - H_{abcd}^F) \rt. \nonu \\ 
&&  \lt.  + 10\ka \sum_i Y^{2F}(i) \la_{abcd} \rt\} \nonu \\
&-& g^4 \lt\{ \lt[ {35\ov3}C_2(G) - {10\ov3}\ka S_2(F) - {11\ov12}S_2(S) \rt]
    \Lambda_{abcd}^S \rt. \nonu \\
&& \ \ \ \ \ \ \ -{3\ov2}\Lambda_{abcd}^{SS} - {5\ov2}A_{abcd}^\la
   - {1\ov2}\bar{A}_{abcd}^\la  \nonu \\ 
&& \ \ \ \ \ \ \  \lt. + 4\ka (B_{abcd}^Y - 10\bar{B}_{abcd}^Y) \rt\} \nonu \\
&+& g^6 \lt\{ \lt[ {161\ov6}C_2(G) - {32\ov3}\ka S_2(F) - {7\ov3}S_2(S) \rt] A_{abcd} \rt. \nonu \\
&& \ \ \ \ \ \ \ \lt. - {15\ov2}A_{abcd}^S + 27A_{abcd}^g \rt\}
\label{sca4}
\ea
where $\Lambda^2(i)$, $H^2(i)$, $\bar{H}^2(i)$ and $Y^{2F}(i)$ are the eigenvalues of the invariants
of $\Lambda_{ab}^2(S), \ H_{ab}^2(S), \ \bar{H}_{ab}^2(S), \ Y_{ab}^{2F}(S)$, respectively.
The other invariants are defined by
\ba
\bar{\Lambda}_{abcd}^3 &=& {1\ov4} \sum_{perms} \la_{abef} \la_{cegh} \la_{dfgh} \\
\bar{\Lambda}_{abcd}^{2Y} &=& {1\ov8} \sum_{perms} Y_2^{fg}(S) \la_{abef} \la_{cdeg} \\
\bar{H}_{abcd}^{\la} &=& {1\ov8}\sum_{perms} \la_{abef} \tr (Y^c Y^{+e} Y^d Y^{+f} \nonu \\
  && \ \ \ \ \ \ \ + Y^{+c} Y^e Y^{+d} Y^f) \\
H_{abcd}^Y &=& \sum_{perms} \tr \left[ Y_2(F) Y^{+a} Y^b Y^{+c} Y^d \right] \\
\bar{H}_{abcd}^Y &=& {1\ov2} \sum_{perms} \tr (Y^e Y^{+a} Y^e Y^{+b} Y^c Y^{+d} \nonu \\
  && \ \ \ \ \ \ \ + Y^{+e} Y^a Y^{+e} Y^b Y^{+c} Y^d ) \\
H_{abcd}^3 &=& {1\ov2} \sum_{perms} \tr (Y^a Y^{+b} Y^e Y^{+c} Y^d Y^{+e}) \\
\bar{\Lambda}_{abcd}^{2S} &=& {1\ov8} \sum_{perms} C_2^{fg}(S) \la_{abef} \la_{cdeg} \\
\Lambda_{abcd}^{2g} &=& {1\ov8} \sum_{perms} \la_{abef} \la_{cdgh} \th_{eg}^A \th_{fh}^A \\
H_{abcd}^S &=& \sum_i C_2(i) H_{abcd} \\
H_{abcd}^F &=& \sum_{perms} \tr \lt[ \{C_2(F), Y^a \}Y^{+b} Y^c Y^{+d} \rt] \\
\Lambda_{abcd}^{SS} &=& \sum_i \lt[ C_2(i) \rt]^2 \la_{abcd} \\
A_{abcd}^\la &=& {1\ov4} \sum_{perms} \la_{abef} \{ \th^A, \th^B \}_{ef} \{\th^A, \th^B\}_{cd} \\
\bar{A}_{abcd}^\la &=& {1\ov4} \sum_{perms} \la_{abef} \{\th^A, \th^B\}_{ce} \{\th^A,\th^B\}_{df} \\
B_{abcd}^Y &=& {1\ov4} \sum_{perms} \{ \th^A, \th^B\}_{ab}
               \tr \lt[ t^{A*} t^{B*} Y^c Y^{+d} \rt. \nonu \\ 
  && \ \ \ \ \ \ \  \lt. + Y^c t^A t^B Y^{+d} \rt] \\
\bar{B}_{abcd}^Y &=& {1\ov4} \sum_{perms} \{ \th^A, \th^B \}_{ab} \tr ( t^{A*} Y^c t^B Y^{+d} ) \\
A_{abcd}^S &=& \sum_i C_2(i) A_{abcd} \\
A_{abcd}^g &=& {1\ov8} f^{ACE} f^{BDE} \sum_{perms} \{\th^A,\th^B\}_{ab} \{\th^C,\th^D\}_{cd}
\ea
In the first term of $B_{abcd}^Y$, $Y^c$ is preceded by $t^{A*} t^{B*}$ instead of $t^A t^B$, since
the fermions are complex.
Similarly in $\bar{B}_{abcd}^Y$, $Y^c$ is preceded by $t^{A*}$ instead of $t^A$.
Since there is only one $t$ factor in this case, this introduces one extra minus sign. 
Therefore, the relative sign between $B_{abcd}^Y$ and $\bar{B}_{abcd}^Y$ in Eq.(\ref{sca4})
is minus while it was plus in \cite{MV3}.
In addition to $H_{ab}^2(S), \ \bar{H}_{ab}^2(S)$, and $Y_{ab}^{2F}(S)$, 
$\bar{H}_{abcd}^Y$ is also re-expressed to be hermitian.

\section{Dimensional parameters}
\subsection{Fermion mass}
The $\beta$-functions of fermion mass can be inferred from those of the Yukawa couplings
by taking the $a$-indices to be dummy.
The trilinear scalar terms start to contribute from two-loops.
The one loop result is
\ba
\beta_{m_f}^{I}  =&&  {1\ov2} \lt[Y_2^+(F) m_f + m_f Y_2(F) \rt] + 2 Y^b m_f^+ Y^b \nonu \\
&&    + \ka Y^b \tr(m_f^+ Y^b + m_f Y^{+b}) \nonu \\ 
&&    - 3 g^2 \{ C_2(F), m_f \}
\ea
The two loop result is
\ba
\beta_{m_f}^{II} &=& 2 Y^c Y^{+b} m_f (Y^{+c} Y^b - Y^{+b} Y^c) \nonu \\
&-& Y^b \lt[Y_2(F) m_f^+ + m_f^+ Y_2^+(F) \rt] Y^b \nonu \\
&-& {1\ov8} \lt[Y^b Y_2(F) Y^{+b} m_f + m_f Y^{+b} Y_2^+(F) Y^b \rt] \nonu \\
&-& 2\ka Y^b Y^{+c} Y^b  \tr (m_f^+ Y^c + m_f Y^{+c})  \nonu \\ 
&-& {3\ov2}\ka Y_2^{bc}(S) (Y^b Y^{+c} m_f + m_f Y^{+c} Y^b ) \nonu \\
&-& {3\ov2}\ka Y^b \tr \lt[ Y_2(F) Y^{+b} m_f + m_f^+ Y_2^+(F) Y^b \rt] \nonu \\
&-& \ka Y^b \tr (Y^c m_f^+ Y^c Y^{+b} + Y^{+c} m_f Y^{+c} Y^b) \nonu \\
&-& 2 h_{bcd} Y^b Y^{+c} Y^d + {1\ov12} h_{cde} \la_{bcde} Y^b \nonu \\
&+& 3 g^2 \{ C_2(F), Y^b m_f^+ Y^b \} + 5 g^2 Y^b \{ C_2(F), m_f^+ \} Y^b \nonu \\
&-& {7\ov4} g^2 [C_2(F) Y_2^+(F) m_f + m_f Y_2(F) C_2(F)] \nonu \\
&-& {1\ov4} g^2 [Y^b C_2(F) Y^{+b} m_f + m_f Y^{+b} C_2(F) Y^b] \nonu \\
&+& 6 g^2 [t^{A*} m_f Y^{+b} t^{A*} Y^b + Y^b t^A Y^{+b} m_f t^A] \nonu \\
&+& 5 \ka g^2 Y^b \tr [C_2(F) (m_f Y^{+b} + Y^b m_f^+)] \nonu \\
&+& 6 g^2 C_2^{bc}(S) Y^b m_f^+ Y^c - {3\ov2} g^4 \{ \lt[ C_2(F) \rt]^2, m_f \} \nonu \\
&+& {9\ov2} g^2 C_2^{bc}(S) (Y^b Y^{+c} m_f + m_f Y^{+c} Y^b) \nonu \\
&+& g^4 \lt[ - {97\ov6}C_2(G) + {10\ov3}\ka S_2(F) \rt. \nonu \\ 
&& \lt. \ \ \ \ \ \ + {11\ov12}S_2(S) \rt] \{ C_2(F), m_f \}
\ea

\subsection{Trilinear scalar couplings}
The $\beta$-functions of trilinear scalar couplings can be inferred 
from those of the quartic couplings by taking one of the four indices to be dummy.
The fermion masses contribute from one-loop.  
One loop result is
\ba
\beta_{h_{abc}}^{I} &=&
\Lambda_{abc}^2 - 8\ka H_{abc} + 2\ka \Lambda_{abc}^Y - 3g^2 \Lambda_{abc}^S
\ea
where the invariants are defined as
\ba
\Lambda_{abc}^2 &=& {1\ov2} \sum_{perms} \la_{abef} h_{efc} \\
H_{abc} &=& {1\ov2} \sum_{perms} \tr(m_f Y^{+a} Y^b Y^{+c} + Y^a m_f^+ Y^b Y^{+c}) \\
\Lambda_{abc}^Y &=& \sum_i Y_2(i) h_{abc} \\
\Lambda_{abc}^S &=& \sum_i C_2(i) h_{abc}
\ea
The two loop result is
\ba
\beta_{h_{abc}}^{II} &=& {1\ov2} \sum_i \Lambda^2(i) h_{abc} 
 -  \bar{\Lambda}_{abc}^3 - 4\ka \bar{\Lambda}_{abc}^{2Y}  \nonu \\ 
&+& \ka \lt\{ 8\bar{H}_{abc}^{\la m} + 8\bar{H}_{abc}^h \rt.  \nonu \\
&& \ \ \ \  \lt.  - \sum_i \lt[ 3H^2(i) + 2\bar{H}^2(i) \rt]  h_{abc} \rt\} \nonu \\
&+& 4\ka(H_{abc}^Y + 2\bar{H}_{abc}^Y + 2H_{abc}^3) \nonu \\
&+& g^2 \lt\{ 2\bar{\Lambda}_{abc}^{2S} - 6\Lambda_{abc}^{2g} + 4\ka(H_{abc}^S - H_{abc}^F)
   \rt. \nonu \\
&& \ \ \ \  \lt.    + 10\ka \sum_i Y^{2F}(i) h_{abc} \rt\} \nonu \\
&-& g^4 \lt\{ \lt[ {35\ov3}C_2(G) - {10\ov3}\ka S_2(F) - {11\ov12}S_2(S) \rt]
    \Lambda_{abc}^S \rt. \nonu \\
&& \ \ \ \ \ \ \ -{3\ov2}\Lambda_{abc}^{SS} - {5\ov2}A_{abc}^\la
   - {1\ov2}\bar{A}_{abc}^\la \nonu \\ 
&& \ \ \ \ \ \ \    \lt. + 4\ka (B_{abc}^Y - 10\bar{B}_{abc}^Y) \rt\}
\ea
where the invariants are defined as
\ba
\bar{\Lambda}_{abc}^3 &=&
{1\ov2} \sum_{perms} (\la_{abef} \la_{cegl} h_{fgl} + \la_{aegl} \la_{bfgl} h_{cef} ) \\
\bar{\Lambda}_{abc}^{2Y} &=&
{1\ov2} \sum_{perms} Y_2^{fg}(S) \la_{abef} h_{ceg} \\
\bar{H}_{abc}^{\la m} &=&
{1\ov8} \sum_{perms} \la_{abef} \tr \left( Y^c Y^{+e} m_f Y^{+f} \rt.  \nonu \\ 
&& \ \ \ \ \ \ \ \  + m_f Y^{+e} Y^c Y^{+f} + Y^{+c} Y^e m_f^+ Y^f \nonu \\
&& \ \ \ \ \ \ \ \  \lt.  + m_f^+ Y^e Y^{+c} Y^f \rt) \\
\bar{H}_{abc}^h &=&
{1\ov4}\sum_{perms} h_{aef} \tr (Y^b Y^{+e} Y^c Y^{+f}  \nonu \\ 
&& + Y^{+b} Y^e Y^{+c} Y^f) \\
H_{abc}^Y &=&
\sum_{perms} \tr \lt[ Y_2(F) ( m_f^+ Y^a Y^{+b} Y^c \right. \nonu \\ 
&&  \ \ \ \ \ \ \ \   + Y^{+a} m_f Y^{+b} Y^c + Y^{+a} Y^b m_f^+ Y^c  \nonu \\ 
&&  \ \ \ \ \ \ \ \   \lt. + Y^{+a} Y^b Y^{+c} m_f) \rt] \\
\bar{H}_{abc}^Y &=&
{1\ov2} \sum_{perms} \tr \lt( Y^e m_f^+ Y^e Y^{+a} Y^b Y^{+c} \rt. \nonu \\ 
&&  \ \ \ \ \ \ \ \   + Y^e Y^{+a} Y^e m_f^+ Y^b Y^{+c} \nonu \\ 
&&  \ \ \ \ \ \ \ \   + Y^e Y^{+a} Y^e Y^{+b} m_f Y^{+c} \nonu \\
&&  \ \ \ \ \ \ \ \   \lt. + Y^e Y^{+a} Y^e Y^{+b} Y^c m_f^+ + h.c \rt) \\
H_{abc}^3 &=&
{1\ov2} \sum_{perms} \tr \left( m_f Y^{+a} Y^e Y^{+b} Y^c Y^{+e} \rt. \nonu \\ 
&&  \ \ \ \ \ \ \ \   + Y^a m_f^+ Y^e Y^{+b} Y^c Y^{+e} \nonu \\ 
&&  \ \ \ \ \ \ \ \   + Y^a Y^{+b} Y^e m_f^+ Y^c Y^{+e} \nonu \\
&&  \ \ \ \ \ \ \ \   + \lt. Y^a Y^{+b} Y^e Y^{+c} m_f Y^{+e} \rt) \\
\bar{\Lambda}_{abc}^{2S} &=&
{1\ov2} \sum_{perms} C_2^{fg}(S) h_{aef} \la_{bceg} \\
\Lambda_{abc}^{2g} &=&
{1\ov2} \sum_{perms} h_{aef} \la_{bcgl} \th_{eg}^A \th_{fl}^A \\
H_{abc}^S &=& \sum_i C_2(i) H_{abc}  \\
H_{abc}^F &=&
\sum_{perms} \tr \lt[ \{C_2(F), m_f\}Y^{+a} Y^b Y^{+c} \rt. \nonu \\
&&  \ \ \ \ \ \ \ \   + \{C_2(F), Y^a\}m_f^+ Y^b Y^{+c} \nonu \\ 
&&  \ \ \ \ \ \ \ \   + \{C_2(F), Y^a\}Y^{+b} m_f Y^{+c} \nonu \\
&&  \ \ \ \ \ \ \ \   + \lt. \{C_2(F), Y^a\}Y^{+b} Y^c m_f \rt] \\
\Lambda_{abc}^{SS} &=& \sum_i \lt[ C_2(i) \rt]^2 h_{abc} \\
A_{abc}^\la &=& {1\ov2} \sum_{perms} h_{aef} \{\th^A,\th^B\}_{ef} \{\th^A,\th^B\}_{bc} \\
\bar{A}_{abc}^\la &=& {1\ov2} \sum_{perms} h_{aef} \{\th^A,\th^B\}_{be} \{\th^A,\th^B\}_{cf} \\
B_{abc}^Y &=& {1\ov4} \sum_{perms} \{\th^A,\th^B\}_{ab}
              \tr \lt( t^{A*} t^{B*} m_f Y^{+c} \rt. \nonu \\
&&  \ \ \ \ \ \ \ \  + m_f t^A t^B Y^{+c} + t^{A*} t^{B*} Y^c m_f^+ \nonu \\ 
&&  \ \ \ \ \ \ \ \  \lt. + Y^c t^A t^B m_f^+ \rt) \\
\bar{B}_{abc}^Y &=& {1\ov4} \sum_{perms} \{\th^A,\th^B\}_{ab}
                    \tr\lt( t^{A*} m_f t^B Y^{+c} \rt. \nonu \\ 
&& \ \ \ \ \ \ \ \ \lt. + t^{A*} Y^c t^B m_f^+ \rt)
\ea

\subsection{Scalar mass}
The $\beta$-functions of scalar masses can also be inferred 
from those of the quartic couplings by taking two of the four indices to be dummy.
From one loop, both fermion masses and trilinear terms contribute.  
One loop result is
\ba
\beta_{m_{ab}^2}^I  &=&
m_{ef}^2 \la_{abef} +  h_{aef} h_{bef} - 4 \ka H_{ab} \nonu \\  
&& - 3g^2 \Lambda_{ab}^S + 2 \ka \Lambda_{ab}^Y 
\ea
where the invariants are defined as
\ba
H_{ab} &=& \tr \lt[ \ \
(Y^a Y^{+b} + Y^b Y^{+a}) m_f m_f^+ \rt. \nonu \\
&& \ \ \ \ + (Y^{+a} Y^b + Y^{+b} Y^a) m_f^+ m_f \nonu \\
&& \ \ \ \ \lt. + Y^a m_f^+ Y^b m_f^+ + m_f Y^{+a} m_f Y^{+b} \rt]  \\
\Lambda_{ab}^S &=& \sum_i C_2(i) m_{ab}^2 \\
\Lambda_{ab}^Y &=& \sum_i Y_2(i) m_{ab}^2
\ea
The two-loop contribution is
\ba
\beta_{m^2_{ab}}^{II} &=& 
{1\ov2} \sum_i \Lambda^2(i) m^2_{ab} - {1\ov2} \bar{\Lambda}_{ab}^3 - 4\ka \bar{\Lambda}_{ab}^{2Y} \nonu \\
&+& \ka \lt\{ 4\bar{H}_{ab}^\la - \sum_i \lt[ 3H^2(i) + 2\bar{H}^2(i) \rt] m_{ab}^2 \rt\} \nonu \\
&+& 2\ka(H_{ab}^Y + 2\bar{H}_{ab}^Y + 2H_{ab}^3) \nonu \\
&+& g^2 \lt\{ 2\bar{\Lambda}_{ab}^{2S} - 6\Lambda_{ab}^{2g} + 2\ka(H_{ab}^S - H_{ab}^F) \rt.  \nonu \\
&+& \lt. 10\ka \sum_i Y^{2F}(i) m_{ab}^2 \rt\} \nonu \\  
&-& g^4 \lt\{ \lt[ {35\ov3}C_2(G) - {10\ov3}\ka S_2(F) - {11\ov12}S_2(S) \rt]
    \Lambda_{ab}^S \rt. \nonu \\
&& \ \ \ \ \ \ \ -{3\ov2}\Lambda_{ab}^{SS} - {5\ov2}A_{ab}^\la
   - {1\ov2}\bar{A}_{ab}^\la \nonu \\  
&& \ \ \ \ \ \ \  \lt. + 2\ka (B_{ab}^Y - 10\bar{B}_{ab}^Y) \rt\}
\ea
where the invariants are defined as
\ba
\bar{\Lambda}_{ab}^3 &=& \la_{abef} h_{egl} h_{fgl} + 2 m_{ef}^2 \la_{aegl} \la_{bfgl} \nonu \\
&& + 2 h_{aef} h_{fgl} \la_{begl} + 2 h_{bef} h_{fgl} \la_{aegl}  \\
\bar{\Lambda}_{ab}^{2Y} &=& Y_2^{fg}(S)(m_{eg}^2 \la_{abef} + h_{aef} h_{beg}) \\
\bar{H}_{ab}^\la &=&
  {1\ov2} \la_{abef} \tr(m_f Y^{+e} m_f Y^{+f} + h.c.) \nonu \\ 
&& \ + m_{ef}^2 \tr(Y^a Y^{+e} Y^b Y^{+f} + h.c.) \nonu \\
&& \ + h_{aef} \tr(Y^b Y^{+e} m_f Y^{+f} + h.c.) \nonu \\
&& \ + h_{bef} \tr(Y^a Y^{+e} m_f Y^{+f} + h.c.) \\
H_{ab}^Y &=&  2\tr \lt[ \{Y_2(F), m_f^+ m_f\} (Y^{+a} Y^b + Y^{+b} Y^a) \rt] \nonu \\
&+&  2\tr \lt[ Y_2(F) Y^{+a} m_f (Y^{+b} m_f + m_f^+ Y^b) \rt. \nonu \\
&& \ + Y_2(F) m_f^+ Y^a (Y^{+b} m_f + m_f^+ Y^b)   \nonu \\
&& \ + Y_2(F) Y^{+b} m_f (Y^{+a} m_f + m_f^+ Y^a) \nonu \\
&& \ \lt. + Y_2(F) m_f^+ Y^b (Y^{+a} m_f + m_f^+ Y^a) \rt] \\
\bar{H}_{ab}^Y &=& \tr \lt[
(Y^e Y^{+a} Y^e Y^{+b} + Y^e Y^{+b} Y^e Y^{+a}) m_f m_f^+ \rt. \nonu \\
&& \ + Y^e m_f^+ Y^e m_f^+ (Y^a Y^{+b} + Y^b Y^{+a}) \nonu \\
&& \ + Y^e Y^{+a} Y^e m_f^+ (Y^b m_f^+ + m_f Y^{+b}) \nonu \\
&& \ + Y^e m_f^+ Y^e Y^{+a} (Y^b m_f^+ + m_f Y^{+b}) \nonu \\
&& \ + Y^e Y^{+b} Y^e m_f^+ (Y^a m_f^+ + m_f Y^{+a}) \nonu \\
&& \ \lt. + Y^e m_f^+ Y^e Y^{+b} (Y^a m_f^+ + m_f Y^{+a}) + h.c. \rt] \\
H_{ab}^3 &=& \tr \lt[ 
(Y^a Y^{+b} + Y^b Y^{+a}) Y^e m_f^+ m_f Y^{+e} \rt. \nonu \\
&& \ + m_f m_f^+ Y^e (Y^{+a} Y^b + Y^{+b} Y^a) Y^{+e} \nonu \\
&& \ + Y^a m_f^+ Y^e (Y^{+b} m_f + m_f^+ Y^b) Y^{+e} \nonu \\
&& \ + m_f Y^{+a} Y^e (Y^{+b} m_f + m_f^+ Y^b) Y^{+e} \nonu \\
&& \ + Y^b m_f^+ Y^e (Y^{+a} m_f + m_f^+ Y^a) Y^{+e} \nonu \\
&& \ \lt. + m_f Y^{+b} Y^e (Y^{+a} m_f + m_f^+ Y^a) Y^{+e} \rt] \\
\bar{\Lambda}_{ab}^{2S} &=& C_2^{fg}(S) \la_{abef} m_{eg}^2 + C_2^{fg}(S) h_{aef} h_{beg} \\
\Lambda_{ab}^{2g} &=& \la_{abef} m_{gl}^2 + h_{aef} h_{bgl} \th_{eg}^A \th_{fl}^A \\
H_{ab}^S &=& \sum_i C_2(i) H_{ab} \\
H_{ab}^F &=& 2 \tr \lt[ \ \
\{C_2(F),Y^a\} Y^{+b} m_f m_f^+ \rt. \nonu \\
&& \ + \{C_2(F),Y^b\} Y^{+a} m_f m_f^+ \nonu \\
&& \ + \{C_2(F),m_f\} m_f^+ (Y^a Y^{+b} + Y^b Y^{+a}) \nonu \\
&& \ + \{C_2(F),Y^a\} m_f^+ (Y^b m_f^+ + m_f Y^{+b}) \nonu \\
&& \ + \{C_2(F),m_f\} Y^{+a} (Y^b m_f^+ + m_f Y^{+b}) \nonu \\
&& \ + \{C_2(F),Y^b\} m_f^+ (Y^a m_f^+ + m_f Y^{+a}) \nonu \\
&& \ \lt. + \{C_2(F),m_f\} Y^{+b} (Y^a m_f^+ + m_f Y^{+a}) \rt] \\
\Lambda_{ab}^{SS} &=& \sum_i \lt[ C_2(i) \rt]^2 m_{ab}^2 \\
A_{ab}^\la &=& m_{ef}^2 \{\th^A,\th^B\}_{ef} \{\th^A,\th^B\}_{ab} \\
\bar{A}_{ab}^\la &=& m_{ef}^2 \{\th^A,\th^B\}_{ae} \{\th^A,\th^B\}_{bf} \\
B_{ab}^Y &=& \{\th^A,\th^B\}_{ab} \tr(t^{A*} t^{B*} m_f m_f^+ + m_f t^A t^B m_f^+) \\
\bar{B}_{ab}^Y &=& \{\th^A,\th^B\}_{ab} \tr(t^{A*} m_f t^B m_f^+)
\ea

\section{Extension to non-simple groups}
So far, the gague group is assumed to be simple.
These results can be extended to semi-simple groups by assigning
appropriate substitution rules, based upon close inspections of relevant Feynman diagrams
\cite{MV1,MV2,MV3}.  

Assume the gauge group is a direct product simple groups,
$G_1 \times \cdots \times G_n$, with corresponding gauge
coupling constants $g_1, \ldots, g_n$.
The substitution rules for the gauge coupling constants are
\ba
g^3 C_2(G) \rightarrow && g_k^3 C_2(G_k) \\
g^3 S_2(R) \rightarrow && g_k^3 S_2^k(R) \\
g^5 [C_2(G)]^2 \rightarrow && g_k^5 [C_2(G_k)]^2 \\
g^5 C_2(G) S_2(R) \rightarrow && g_k^5 C_2(G_k) S_2^k(R) \\
g^5 C_2(R) S_2(R) \rightarrow && \sum_l g_k^3 g_l^2 C_2^l(R) S_2^k(R)
\ea
here and hereafter, $k$ and $l$ are subgroup indices, $R$ can be either $S$ or $F$. 

For other $\beta$ and $\gamma$ functions, we first have the following general
substitution rules 
\ba
g^2 C_2(R) \rightarrow && \sum_k g_k^2 C_2^k(R) \\
g^4 C_2(G) C_2(R) \rightarrow && \sum_k g_k^4 C_2(G_k) C_2^k(R) \\
g^4 S_2(R) C_2(R') \rightarrow && \sum_k g_k^4 S_2^k(R) C_2^k(R') \\
g^4 C_2(R) C_2(R') \rightarrow && \sum_{k,l} g_k^2 g_l^2 C_2^k(R) C_2^l(R')
\ea

In $H_{2t}^a, \ B_{abcd}^Y, \ B_{abc}^Y, \  B_{ab}^Y, \  \bar{B}_{abcd}^Y, \ \bar{B}_{abc}^Y, \ \bar{B}_{ab}^Y$, 
the substitution rules are 
\ba
\th^A \rightarrow \th_k^A, && \th^B \rightarrow \th_l^B \nonu \\
t^A \rightarrow t_k^A, && t^B \rightarrow t_l^B \nonu \\
(t^{A*} \rightarrow t_k^{A*}, &&  t^{B*} \rightarrow t_l^{B*}) \nonu \\
g^4 \rightarrow && g_k^2 g_l^2
\ea
For example, $g^4 B_{abcd}^Y$ is substituted by
\be
{1\ov4} \sum_{k,l} g_k^2 g_l^2 \sum_{perms} \{\th_k^A,\th_l^B\}_{ab}
\tr (t_k^{A*} t_l^{B*} Y^c Y^{+d} + Y^c t_k^A t_l^B Y^{+d})
\ee

In quartic coupling, trilinear coupling and scalar mass-square terms, 
further substitution rules are needed. Introduce a new tensor 
\be 
\Lambda_{ab,cd}=(\th^A)_{ac} (\th^A)_{bd} 
\ee
so the gauge invariants $A_{abcd}$ can be rewritten as 
\be 
 A_{abcd} = {1\ov4} \sum_{perms}
(\Lambda_{ac,ef} \Lambda_{ef,bd} + \Lambda_{ae,fd} \Lambda_{eb,cf}).
\ee
The substitution rule for $\Lambda_{ab,cd}$ is 
\be 
g^2 \Lambda_{ab,cd} \rightarrow \sum_k g_k^2 \Lambda_{ab,cd}^k 
\ee 
Thus $g^4 A_{abcd}$ can be substituted by 
\be
{1\ov4} \sum_{k,l} g_k^2 g_l^2
\sum_{perms}(\Lambda_{ac,ef}^k \Lambda_{ef,bd}^l + \Lambda_{ae,fd}^k \Lambda_{eb,cf}^l)
\ee
$g^6 S_2(R) A_{abcd}$ by 
\be 
{1\ov4} \sum_{k,l} g_k^4 g_l^2 S_2^k(R)
\sum_{perms} (\Lambda_{ac,ef}^k \Lambda_{ef,bd}^l + \Lambda_{ae,fd}^k \Lambda_{eb,cf}^l) 
\ee
$g^6 C_2(G) A_{abcd}$ by
\be 
{1\ov4} \sum_{k,l} g_k^4 g_l^2 C_2(G_k)
\sum_{perms} (\Lambda_{ac,ef}^k \Lambda_{ef,bd}^l + \Lambda_{ae,fd}^k \Lambda_{eb,cf}^l)
\ee
and $g^6 A_{abcd}^S$ by
\ba
 \sum_k \sum_i g_k^2 C_2^k(i) {1\ov4} \sum_{l,m} g_l^2 g_m^2  \nonu \\
\times \sum_{perms} (\Lambda_{ac,ef}^l \Lambda_{ef,bd}^m &+& \Lambda_{ae,fd}^l \Lambda_{eb,cf}^m)
\ea
Finally, one has
\be
g^6 A_{abcd}^g \rightarrow \sum_{k} g_k^6 A_{abcd}^g(k)
\ee

\section{Conclusions}
We have now presented the complete set of 
two-loop renormalization group equations in general gauge theories.
This included the $\beta$-functions of parameters with and without a mass
dimension.
We have so far restricted the gauge groups to be semi-simple.
If the gauge group contains more than one $U(1)$ group, the situation is subtle.
If there are no kinetic mixings between the $U(1)$ groups after renormalization,
these results can applied by straightforward extension.
If there are kinetic mixings, 
these results have to be applied with caution. 
In general, modifications will be warranted \cite{luo2}.
In the case of the SM where one has only one $U(1)$ group, 
one readily reproduces the results in \cite{luo}.

\begin{acknowledgments} 
We thank S. Martin for valuable discussions.
The work was supported by a Fund for Trans-Century Talents,
CNSF-90103009  and CNSF-10047005. 
\end{acknowledgments}


\begin{thebibliography}{10}
\bibitem{pdg}
  K. Hagiwara et al., \prt{66}{010001} (2002).
\bibitem{erler}
  J. Erler and P. Langacker, p.98 in \cite{pdg}.
\bibitem{GUT1}
  P. Langacker and M. Luo, \prt{44}{817} (1991).
\bibitem{GUT2}
  U. Amaldi, W. de Boer and H. F\"{u}rstenau, \plb{260}{447} (1991).
\bibitem{GUT3}
  J. Ellis, S. Kelley and D. Nanopoulos, \plb{260}{131} (1991). 
\bibitem{4loops}
  T. van Ritbergen, J. Vermaseren and S. Larin, \plb{400}{379} (1997).
\bibitem{MV1}
M.E. Machacek and M.T. Vaughn, \npb{222}{83} (1983)
\bibitem{MV2}
M.E. Machacek and M.T. Vaughn, \npb{236}{221} (1984)
\bibitem{MV3}
M.E. Machacek and M.T. Vaughn, \npb{249}{70} (1985)
\bibitem{martin}
  S. Martin and M. Vaughn, \prt{50}{2282} (1994). 
\bibitem{luo}
M. Luo and Y. Xiao, \pru{90}{011601}(2003)(hep-ph/0207271). 
\bibitem{luo2}
M. Luo and Y. Xiao, \plb{555}{279}(2003)(hep-ph/0212152).
\end{thebibliography}
\end{document}